\documentclass [preprint, prd, aps, nofootinbib]{revtex4}

\begin{document}
\draft
\title{\large \bf Holographic Foam, Dark Energy and Infinite Statistics}

\author{
Y. Jack Ng\footnote{{\it E-mail address}:
yjng@physics.unc.edu}}
\address{Institute of Field Physics, Department
of Physics and Astronomy, University of North Carolina, Chapel
Hill, NC 27599, USA\\}

\bigskip

\begin{abstract}

Quantum fluctuations of spacetime give rise to quantum foam, and
black hole physics dictates that the foam is of holographic 
type.  Applied to cosmology, the holographic model requires the existence of 
dark energy which, we argue, is composed 
of an enormous number of inert ``particles''
of extremely long wavelength.  
These ``particles" 
necessarily obey
infinite statistics in which all representations of the particle 
permutation group can occur. 
For every boson or fermion in the present observable universe 
there could be $\sim 10^{31}$ such ``particles". 
We also discuss the compatibility between the holographic principle 
and infinite statistics.

\bigskip

PACS numbers: 04.60.-m, 95.36.+x, 05.30.-d

\end{abstract}

\maketitle

\section{Introduction}

According to folklore,
there are two 
kinds of statistics: Fermi-Dirac statistics for identical particles of 
half-integral spin and Bose-Einstein statistics for identical particles of 
integral spin.  (There 
are also generalizations of these statistics known as para-Fermi 
and para-Bose statistics. \cite{green})  But it is far less well-known
that there
is a third kind of particle statistics,
known as infinite statistics
\cite{DHR,govorkov,greenberg},
that is consistent with the general 
principles of quantum field theory. 
A collection of particles obeying infinite statistics can be in any 
representation of the particle permutation group: compare this with
the rule that fermions (bosons) can only be in a totally antisymmetric 
(symmetric) state.  While there are plenty of examples of fermions and bosons,
there is no {\it empirical evidence} for particles of infinite statistics --- 
until now perhaps.  In this Letter, we will argue that actually  
infinite statistics should be the most familiar kind of statistics.  For 
every observed fermion or 
boson,\footnote{Not including the degrees of freedom behind black hole horizons, 
nor the entropy of gravitons and dark matter.} 
there could be as many as $\sim 10^{31}$ ``particles'' 
obeying infinite statistics in the observable universe in the present cosmic
era.  

We will apply our argument to the cosmology derived from 
the so-called holographic
model of spacetime foam.  The outline of this Letter is as follows.
In Section II, partly for completeness, 
we explain the logic behind the holographic quantum foam.  
In Section III, we show that the holographic model, 
applied to cosmology, predicts that the cosmic energy is of critical density,
and the cosmic entropy is the maximum allowed by the holographic 
principle.  We also discuss the random-walk model which predicts a 
coarser spatial resolution in the mapping of spacetime geometry than the
holographic model.  Existing archived data on quasars from the Hubble 
Space Telescope can be used to rule out the random-walk model the demise 
of which, coupled with the fact (see below) that ordinary matter composed 
of fermions and/or bosons maps out spacetime only to the accuracy
corresponding to the random-walk model, can be used to infer the existence 
of unconventional energy and/or matter (independent of recent 
cosmological observations).  The main part of our argument is given in
Section IV: there we show that, in the 
framework of holographic foam cosmology,
positivity of entropy requires
the ``particles'' (or bits) constituting dark energy to 
obey infinite statistics; we also discuss the compatibility between
the holographic principle and infinite statistics. 
We address some issues facing holographic foam cosmology and give 
a summary in the final section.

\section{Holographic Quantum Foam}

Conceivably spacetime, like everything else,
is subject to quantum fluctuations.  As a result, spacetime is ``foamy" at
small scales,
\cite{Wheeler} giving rise to a microscopic structure of spacetime known as
quantum foam, also known as spacetime foam, and
entailing an intrinsic limitation $\delta l$ to the accuracy
with which one can measure a distance $l$.  In principle, $\delta l$ can
depend on both
$l$ and the Planck length $l_P = \sqrt{\hbar G/c^3} $, the intrinsic scale in
quantum gravity,
and hence can be written as
$\delta l \gtrsim l^{1 - \alpha} l_P^{\alpha}$, \footnote{We find it reasonable
and consistent to assume that $\delta l$ depends on only $l$ and $l_P$.}
with $\alpha
\sim 1$ parametrizing the various spacetime foam models.  (For related effects
of quantum fluctuations of spacetime geometry, see Ref.\cite{Ford}.)
In what follows, we will advocate the so-called
holographic model corresponding to $\alpha = 2/3$, but we will also consider
the (random walk) model with $\alpha = 1/2$ for comparison.  

Let us first
derive the holographic model \cite{wigner} by using an argument
based on quantum computation \cite{llo04}. 
Since quantum fluctuations of spacetime 
manifest themselves in the form of uncertainties in the geometry of
spacetime, the structure of spacetime foam can be inferred from the
accuracy with which we can measure that geometry.  Let us
consider a spherical volume of
radius $l$ over the amount of time $T = 2l/c$ it takes light to cross the
volume.  One way to map out the geometry of this spacetime region 
is to fill the space with clocks, exchanging
signals with other clocks and measuring the signals' times of arrival.
This process of mapping the geometry is a sort of computation; 
\footnote{Later, we will extend this process of mapping the geometry to
the entire universe.
For the readers who find the idea of treating the universe as a 
computer unpalatable, we should mention that there are other ways to derive 
the holographic model; see Ref. \cite{wigner}.}
hence
the total number of operations (the ticking of the clocks and
the measurement of signals etc) is bounded by the Margolus-Levitin
theorem\cite{Lloyd}
in quantum computation,
which stipulates that the rate of operations for any computer
cannot exceed the amount of energy $E$ that is available for computation
divided by $\pi \hbar/2$.   A total mass $M$ of clocks then
yields, via the Margolus-Levitin theorem, the bound on the total number of
operations given by $(2 M c^2 / \pi \hbar) \times 2 l/c$.  But to prevent
the clocks from collapsing into a
black hole, $M$ must be less than $l c^2 /2 G$.  Together, these
two limits imply that the total number of operations that can occur in a
spatial volume of radius $l$ for a time period $2 l/c$ is no greater than
$\sim (l/l_P)^2 $.  (Here and henceforth we neglect numerical factors
of order unity, set $c=1=\hbar$ and will also set the Boltzmann constant 
equal to 1.)
To maximize spatial resolution, each clock must tick
only once during the entire time period.  The operations
partition the spacetime volume into ``cells", and, on the average, each cell
occupies a spatial volume no less than $ \sim l^3 / ( l^2 / l_P^2) =
l l_P^2 $, yielding an average separation between neighhoring cells
no less than $l^{1/3} l_P^{2/3}$.  This spatial separation
is a measure of the uncertainty in the geometry of the spacetime volume, and 
hence can be interpreted as yielding an average minimum uncertainty 
\footnote {Note that this result is not inconsistent with that found in
\cite{hsu}.} in the measurement of
a distance $l$ given by $\delta l \gtrsim l^{1/3} l_P^{2/3}$.

Parenthetically we can now understand why this quantum foam model has come to be known
as the holographic model.
Since, on the average, each cell occupies a spatial volume of $l l_P^2$,
a spatial region of size $l$ can contain no more than $l^3/(l l_P^2) =
(l/l_P)^2$ cells.  Thus this model
corresponds to the case of
maximum number of bits of information $l^2 /l_P^2$
in a spatial region of size $l$, that is
allowed by the holographic principle \cite{wbhts}. 

It will prove to be useful to compare the holographic model 
in the mapping of the geometry of spacetime
with the one that corresponds to spreading the spacetime cells uniformly
in both space and time.  For the latter case, each cell has
the size of $(l^2 l_P^2)^{1/4} =
l^{1/2} l_P^{1/2}$ both spatially and temporally so that each clock ticks
once in the time it takes to communicate with a neighboring clock.  Since
the dependence on $l^{1/2}$ is the hallmark of a random-walk fluctuation,
this quantum foam model corresponding to  $\delta l \gtrsim
(l l_P)^{1/2}$ is called the random-walk model \cite{AC}.
Compared to the holographic model, the random-walk model predicts a
coarser spatial resolution, i.e., a larger distance fluctuation,
in the mapping of spacetime geometry.  It
also yields a smaller bound on the information content in a spatial
region, viz., $(l/l_p)^2 / (l/l_P)^{1/2} = (l^2 / l_P^2)^{3/4} =
(l/l_P)^{3/2}$ bits.

One remark is in order.  The minimum $\delta l$ just found 
for the holographic model corresponds
to the case of maximum energy density $\rho \sim (l l_P)^{-2}$ for the
region not
to collapse into a black hole.  Hence the holographic model, 
in contrast to the random-walk model 
\footnote{The random-walk model (corresponding to $\delta l \gtrsim (l l_P)^{1/2}$)
does not require the maximum energy density because
the clocks can tick less frequently than once in the amount of time
$(ll_P)^{1/2}$.}
and other models, requires, for its
consistency, the energy density to have the critical value.

\section{Dark Energy/Matter}

The Planck length $l_P \sim 10^{-33}$ cm is so short that we need an
astronomical (even cosmological) distance $l$ for its fluctuation $\delta
l$ to be detectable.
Let us consider light (with wavelength $\lambda$)
from distant quasars or bright active galactic nuclei. \cite{lie03,NCvD}
Due to quantum fluctuations of spacetime, the wavefront, while planar,
is itself ``foamy", having random fluctuations in phase \cite{NCvD} 
given by $\Delta \phi \sim 2 \pi \delta l / \lambda$ as well as in
the direction \footnote{For this result, we assume 
comparable fluctuations in both the longitudinal and  
transverse components of the wave vector due to spatial isotropy.}
of the wave vector \cite{CNvD} given by $\Delta \phi / 2 \pi  $.
In effect, spacetime foam creates a
``seeing disk" whose angular diameter is $\sim \Delta \phi /2 \pi  $.  For
an interferometer with baseline length $D$, this means that dispersion will
be seen as a spread in
the angular size of a distant point source, causing a reduction in the
fringe
visibility when $\Delta \phi / 2 \pi \sim \lambda / D  $. 

Now we can use existing archived high-resolution data on 
quasars or ultra-bright active galactic nuclei
from the Hubble Space Telescope to test the quantum foam 
models. \cite{CNvD}  Consider the case
of PKS1413+135 \cite{per02}, an AGN for which the redshift is 
$z = 0.2467$.  With $l \approx 1.2$ Gpc and $\lambda = 1.6 \mu$m,
we \cite{NCvD} find $\Delta \phi \sim 10 \times 2 \pi$ and
$10^{-9} \times 2 \pi$ for the random-walk model and
the holographic model of spacetime foam respectively.
With $D = 2.4$ m for HST, we expect to detect halos
if $\Delta \phi \sim 10^{-6} \times 2 \pi$.
Thus, the HST image only fails to test the holographic model by
approximately 3 orders of magnitude.

However, the absence of a spacetime foam induced halo structure in the
HST image of PKS1413+135 rules
out convincingly the random-walk model.  (In fact, the scaling relation
discussed above indicates
that all spacetime foam models with $\alpha \lesssim 0.6$ are ruled
out by this HST observation.)  This result
has profound implications for cosmology. \cite{llo04,CNvD,Arzano}  To wit,
from the observed cosmic critical density in the present era
(consistent with the prediction of the cosmology inspired by the 
holographic model of quantum foam) we deduce that
$\rho \sim H_0^2/G \sim (R_H l_P)^{-2}$, 
where $H_0$ and $R_H$ are the present Hubble parameter 
and Hubble radius of the observable universe respectively.
Treating the whole universe as a computer\cite{llo02, llo04}, one can
apply the Margolus-Levitin theorem to conclude that the universe
\footnote{Note that the total energy of the universe is increasing;
this is due
to the fact that total amount of energy/matter within the horizon is 
growing with time, as more energy/matter enter the horizon.}
computes at a rate $\nu$ up to $\rho R_H^3 \sim R_H l_P^{-2}$
for a total of $(R_H/l_P)^2$
operations during its lifetime so far.
If all the information of this huge computer is stored in ordinary
matter, we can apply standard methods of statistical mechanics
to find that the total number $I$ of bits is $(R_H^2/l_P^2)^{3/4} =
(R_H/l_P)^{3/2} \sim 10^{92}$.
Then each bit flips once in the amount of time given by
$I/\nu \sim (R_H l_P)^{1/2}$. 
On the other hand, the average separation of neighboring bits is
$(R_H^3/I)^{1/3} \sim (R_H l_P)^{1/2}$.
Hence, the time 
to communicate with neighboring bits is equal to the time for each
bit to flip once.  It follows that the accuracy to which ordinary
matter maps out the geometry of spacetime corresponds exactly to
the case of events
spread out uniformly in space and time discussed above for the case
of the random-walk model of spacetime foam.  In other words,
ordinary matter only contains an amount of
information dense enough to map out spacetime at a level consistent with
the random-walk model.  
Observationally ruling out the random-walk model 
suggests that there must be other kinds of matter/energy with which the
universe can map out its spacetime geometry to a finer spatial accuracy
than is possible with the use of conventional matter.  This line of
reasoning
then strongly hints at the existence of dark energy/matter, independent of
the evidence from recent cosmological observations \cite{SNa}.

Moreover, the fact that our universe is observed to be at or very close to
its critical energy density $\rho \sim (H/l_P)^2 \sim (R_H l_P)^{-2}$
must be taken as solid albeit indirect evidence
in favor of the holographic model \cite{FandS}
because, as aforementioned, this model is the only model 
that requires the energy density to have the critical value.  
The holographic model also predicts
a huge number of degrees of freedom for the universe in the present era,
with the cosmic entropy given by \cite{Arzano} 
$I \sim (R_H /l_P)^2 \sim 10^{123}$.
Hence the average energy carried by each bit or ``particle" is
$\rho R_H^3/I \sim R_H^{-1}$.  It is now natural to identify these 
``particles" of unconventional energy/matter 
of extremely long wavelength as constituents of 
dark energy. 
Since altogether $\sim (R_H / l_P)^2$ 
operations have been performed with $\sim (R_H / l_P)^2$ bits, we note, 
for later discussion, that the overwhelming
majority of the bits have had time to flip only of order one time over
the course of cosmic history.  In other words, each ``particle" has 
had only of order one interaction.  The inertness of these ``particles" 
may explain why dark energy is dark.

\section{Infinite Statistics}

What is the overriding difference between conventional matter and
unconventional energy/matter (i.e., dark energy and perhaps also dark matter)?
To find that out, let us first
consider a perfect gas of $N$ particles obeying Boltzmann statistics
(which, rigorously speaking, is not a physical statistics but is still 
a useful statistics to work with) 
at temperature $T$ in a volume $V$.  For the problem 
\footnote{As the lowest-order approximation, let us neglect the 
contributions from matter 
to the cosmic energy density.  Then it can be shown that the 
Friedmann equations for $\rho \sim H^2 /G$ are solved 
by $H \propto 1/a$ and $a \propto t$ with pressure $p \sim -\rho /3$,
where $a(t)$ is the cosmic scale factor.}
at hand, we take $V \sim R_H^3$, $T \sim R_H^{-1}$, 
and very roughly $N \sim (R_H/ l_P)^2$.  
A standard calculation (for the relativistic case) 
yields the partition function 
$Z_N = (N!)^{-1} (V / \lambda^3)^N$, where
$\lambda = (\pi)^{2/3} /T$.
With the free energy given by
$F = -T ln Z_N = -N T [ ln (V/ N \lambda^3) + 1]$,
we get, for the entropy of the system,
\begin{equation}
S = - ( \partial F / \partial T)_{V,N} = N [ln (V / N \lambda^3) + 5/2].
\label{entropy1}
\end{equation}
For the non-relativistic case with the
effective mass $m \sim R_H^{-1}$ (coming from some sort of potential 
with which we are not going to concern ourselves),
the only changes in 
the above expressions are given by the substitution 
$\lambda \longrightarrow (2 \pi / mT)^{1/2}$.
With $m \sim T \sim R_H^{-1}$, 
there is no significant
qualitative difference between the non-relativistic and
relativistic cases.  

The important point to note is that, since $V \sim \lambda^3$, 
the entropy $S$ in 
Eq. (\ref{entropy1}) becomes nonsensically negative unless $ N \sim 1$ 
which is equally nonsensical because $N$ should not be too different from
$(R_H/l_P)^2 \gg 1$.
Intentionally we have calculated the entropy by employing the 
familiar Boltzmann statistics (with the correct Boltzmann counting factor),
only to arrive at a contradictory result.  
But now the solution to this contradiction
is pretty obvious: the $N$ inside the log in Eq. 
(\ref{entropy1}) somehow must be absent.  Then $ S \sim N
\sim (R_H/l_P)^2$ without $N$ being small (of order 1) and S is non-negative 
as physically required.  That is the case if the ``particles" are 
distinguishable and nonidentical!  For in that case, the Gibbs $1/N!$ 
factor is absent from the partition function $Z_N$, and the entropy 
becomes
\begin{equation}
S = N[ln (V/ \lambda^3) + 3/2].
\label{entropy2}
\end{equation}
We can add that, with or without the Gibbs factor, the internal energy is 
given by $U = F + TS = (3/2) NT$.

Now the only known consistent statistics in greater than two space dimensions
without the Gibbs factor 
\footnote{Recall that the Fermi statistics and Bose statistics give
similar results as
the conventional Boltzmann statistics at high temperature.}
is infinite statistics 
(sometimes called
``quantum Boltzmann statistics") \cite{DHR,govorkov,greenberg}.  Thus we have
shown that the ``particles" constituting dark energy obey infinite statistics,
instead of the familiar Fermi or Bose statistics.
What is infinite statistics?  Succinctly, a Fock 
realization of infinite statistics is provided by a $q$ deformation of 
the commutation relations of the oscillators:
$a_k a^{\dagger}_l - q a^{\dagger}_l a_k = \delta_{kl}$ with $q$ between -1 and 1 
(the case $q = \pm 1$ corresponds to bosons or fermions).  States are
built by acting on a vacuum which is annihilated by $a_k$.  Two states
obtained by acting with the $N$ oscillators in different orders are 
orthogonal.  It follows that the states may be in any representation
of the permutation group.  The statistical mechanics of particles 
obeying infinite statistics can be obtained in a way similar to  
Boltzmann statistics, with the crucial difference that the Gibbs
$1/N!$ factor is absent for the former.  Infinite statistics can be
thought of as corresponding to the statistics of identical particles 
with an infinite number of internal degrees of freedom, which is
equivalent to the statistics of nonidentical particles since they are
distinguishable by their internal states.  

Infinite statistics appears to have one ``defect": a theory of 
particles obeying infinite statistics cannot be local \cite{fredenhagen,
greenberg}.  The expressions for the number operator, Hamiltonian, etc.,
are both nonlocal and nonpolynomial in the field operators.  The lack of
locality may make it difficult to formulate a relativistic verion of the
theory; but it appears that a non-relativistic theory can be developed.
Lacking locality also means that the familiar spin-statistics 
relation is no longer valid for particles obeying infinite statistics; hence
they can have any spin.  Remarkably, the TCP theorem and cluster
decomposition have been shown to hold despite the lack of locality. 
\cite{greenberg}

Actually the lack of locality for theories of infinite statistics may 
have a silver lining.  According to the holographic principle, the
number of degrees of freedom in a region of space is bounded not by
the volume but by the surrounding surface.  This suggests that the
physical degrees of freedom are not independent but, considered
at the Planck scale, they must be infinitely correlated, with the result 
that the spacetime location of an event may lose its invariant 
significance.  Since the holographic principle is believed to be
an important ingredient in the formulation of quantum gravity,
the lack of locality for theories of infinite statistics may not be 
a defect; it can actually be a virtue.  Perhaps it is this lack of
locality that makes it easier to incorporate gravitational 
interactions in the theory.  Quantum gravity and infinite statistics 
appear to fit together nicely.  This may be the reason why (charged,
extremal) black holes appear to obey infinite statistics. 
\cite{stromvolo}  
Indirectly this may also explain why the holographic foam model has
use for infinite statistics as we have just shown.

\section{Discussion}

We have considered a perfect gas consisting of 
``particles" of extremely long wavelength,
obeying Boltzmann statistics (first in the conventional, then 
in the quantum version) in the Universe at temperature $T$.  But we have
seen that those ``particles" have had interactions only 
of order one time on the average during the entire cosmic history.
A question can be raised as to whether such an inert gas can come to 
thermal equilibrium at any well defined temperature.  We do not have
a good answer; but the fact that all these ``particles", though
extremely inert, have a wavelength comparable to the observable Hubble 
radius may mean that they overlap significantly, 
\footnote{Thus these ``particles'' provide a spatially 
uniform energy density, like a time-dependent cosmological constant.  But 
in a way, this type of models is preferrable to the cosmological constant
because it may be easier to understand a zero cosmological constant 
(perhaps due to a certain not-yet-known symmetry) than an
exceedingly small (but non-zero) cosmological constant.  We also find it
amusing to recall that earlier cosmic epochs are associated with 
$\rho \propto a^{-4}$ (radiation-dominated) and (followed by)
$\rho \propto a^{-3}$ (matter-dominated).  If the holographic foam cosmology
is correct, these epochs are now succeeded by the dark-energy-dominated
era with $\rho \propto a^{-2}$.}
and accordingly can perhaps share a common temperature.  

Another question concerns the sign of the pressure for this gas 
and whether it is sufficiently negative to accelerate 
the expansion of the present Universe as has been observed.  Indeed
the pressure for such a gas can be easily shown to be $P = (2/3) U/V$
and is blatantly positive.  But this calculation is based on the
simplifying assumption that the gas is perfect.  Such a treatment may
be sufficient for estimating the entropy, but it is obviously inadequate
to give the correct pressure.  After all, as shown above, each
``particle" has an energy comparable to $R_H^{-1}$.  Such
long-wavelength bits or ``particles'' carry negligible kinetic energy.
Since pressure (energy density) is given by kinetic energy minus (plus)
potential energy, a negligible kinetic energy means that
the pressure of the unconventional energy/matter is roughly minus its
energy density, plausibly leading to accelerating cosmic 
expansion.
\footnote{As noted above, for cosmic energy density $\rho \sim H^2/G$, the
equation of state is given by $p \sim - \rho /3$.
To have $p \sim - \rho$, one may need to
take into account, for instance, the interactions between dark energy and 
matter.  See, e.g.,
\cite{negativep}.  At this point, we simply assume that the full dynamics can
generate a sufficiently negative pressure to yield accelerating cosmic 
expansion as observed.  It is also possible that the thermodynamics of
infinite statistics is more complicated than we realize.  Further study is 
warranted.}  This scenario is very similar to that
of quintessence \cite{quint}, but it has its origin in local small scale physics
-- specifically, the holographic quantum foam!

Finally, is there any useful phenomenology that we can predict or use to 
explicity check whether dark energy (and perhaps even dark matter)
is composed of particles obeying infinite statistics?  Since all those
``particles" are so inert, we do not foresee any useful desktop 
experiments forthcoming soon that can shed light on the phenomenology of
dark energy,
a safer bet would be on cosmological
observations (e.g., in connection with the scale-invariance of density
fluctuations \cite{smolin}).  Further study is warranted.

In summary, according to holographic foam cosmology, 
(1) the cosmic energy density takes on the critical value, (2) 
dark energy/matter exists, and (3) the cosmic entropy is the maximum
allowed by the holographic principle.  This scenario may lead to 
cosmic accelerating expansion in the present cosmic era, and interestingly 
it suggests that dark energy is composed of $\sim 10^{123}$ 
extremely cold, inert, and long-wavelength ``particles".
Furthermore we have shown that these ``particles'' 
necessarily obey infinite statistics.  
By a staggering factor of $\sim 10^{123 - 92} =  10^{31}$,
these ``particles" appear to far outnumber 
particles of the familiar Bose and Fermi statistics that we are 
all made of.  Indeed we may be quite insignificant in the cosmic 
grand scheme.
This is a most humbling realization.

\bigskip

{\bf Note added:} After this work was posted on the arXiv (gr-qc/0703096),
we learned of a recent paper \cite{minic} by V.~Jejjala, M.~Kavic 
and D.~Minic on a similar subject.  
In the framework of M-theory, these authors 
argue that dark energy has a fine structure compatible with
infinite statistics.  We also learned of another recent paper \cite{pullin} in
which, in the framework of loop quantum gravity, 
R.~Gambini and J.~Pullin derive, from first principles, the 
fundamental limits on the measurements of space and time and the ultimate 
limits of computability, and they also show the consistency
of these limits with holography.  Their work lends support to the various 
arguments and results found in Ref. \cite{wigner} and \cite{llo04} and 
presented in this paper.

\smallskip

\section*{Acknowledgments}
This work was supported in part by the US Department of Energy and the
Bahnson Fund of the University of North Carolina.  I thank O.~W. 
Greenberg for a brief but useful tutorial on infinite statistics during
the 2005 Miami Conference, and T.~W. Kephart, L.~Mersini, 
R.~Rohm and M.~Arzano
for useful discussions.

\end{document}